# Common Radial Velocity vs. Rare Microlensing: Difficulties and Futures[†]

*Karan Molaverdikhani - K.Molaverdi@Gmail.com*
[†]*PrePublished Ver.*

**Abstract**

In this paper, effective factors for success of Microlensing and Radial Velocity methods were choose. A semi-Delphi process applied on the factors to evaluating them and finding the most important factors for present situation of ML and RV, with help from about 100 experts, in or related exoplanets detection. I found the public definition on "success of exoplanets detection methods" is not correct and we should change it, as some experts did it, in the form of fundamental questions in planetary science. Also, the views of "Special Experts" are different from other experts that help us to choose the right way in evaluating. The next step was choosing the best strategy for future and finally, from SWOT landscape and with a new objective of ML method (New Game Board Strategy) I suggested four critical future strategies for completing current strategic directions.

## Introduction

The number of discovered Extrasolar Planets (ESPs) until 10 Feb 2010 was 429 and about 69 unconfirmed ESPs. But just 10 planets confirmed from 25 potential planets detected by Microlensing (ML) Method (Dominik 2009, Gaudi 2009 unpublished). On the other hand and at the same time, we know about 330 planet that detected by Radial Velocity (RV)[1]. You can see the fraction of the most important detection methods in the Figure-1.

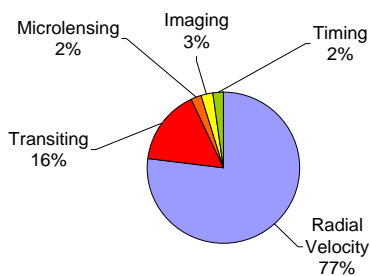

**Figure 1 - Role of exoplanets' detection methods**

Just one ESP detected by astrometry until now and this method used for some previously discovered planets (Pravdo, 2009). The most important advantage of the astrometric method is that it is most sensitive to planets with large orbits. It is good, but it can stand just in long term discovery missions.

The Direct Imaging method is relatively new and scientists hadn't believed in this method until they found the first planet (Fomalhaut-b) in 13 Nov 2008 (Kalas Paul, 2008). But we expect to find about 100 hidden exoplanets just in Hubble's archived pictures since 11 years ago

(O'Neill, 2009). It can help a rapid growth of ESPs in this method, just in next couple of years.

We heard about the Kepler Spacecraft that was launched on March 7, 2009. It continuously monitor the brightness of over 100,000 stars in a fixed field of view and scientists expect to find near 2000 planet in the next 3.5 years[2].

What about Microlensing? The first question is: "Why the number of detected ESPs by ML is very low in comparison to the RV?" And the second question is: "What is the future of ML in ESPs detection?"

For answering to these questions, I used combination of semi-Delphi, AHP and SWOT methods. The first step is describing and evaluating the present situation of ML vs. RV.

Also, this survey is based on "strategic management" not "technology management" and as a result, there isn't any detailed technologically analysis or suggestions. We need separate work(s) with more time and more Delphi rounds to finding critical technologies and scheduling future infrastructure missions.

In many scientific and strategic papers about ESPs detection, target and main goal is "Improvement of Exoplanets Detection" for answering some fundamental questions in planetary science. But this paper wants to show an image of exoplanets' detection competition environment between different techniques with "Development of Microlensing" approach instead "Improvement of Exoplanets Detection". However, development of a technique is helpful for improvement of ESPs detection, too.

---

[1] The Extrasolar Planets Encyclopaedia

[2] Kepler Mission Fact Sheet

## Detection Difficulties and Factors' Effectiveness

The first step is finding and evaluating effective factors on the detection method processes. In Table-1 you can see 19 factors that chose from ML process that potentially can be the reasons of ML rare results vs. RV common results (Gaudi 2009, Schneider 2009, Lawson 2009-2004, Lunine 2008, Ergenzinger 2008, Dominik 2008, Cesarsky 2005 and PPARC 2005).

These factors cover the technical and the non-technical parts of ML method, such as present searching strategy, limited available telescopes or human resources.

The Delphi technique was developed by the RAND Corporation in the late 1960's as a forecasting methodology. And I used a semi-Delphi method that based on a systematic, interactive forecasting survey on a panel of experts. The Delphi method converge the experts answers in two or more rounds (Rowe and Wright 1999). Because of the experts' time limitation and their intense schedule I changed the Delphi process to a modified one.

A semi-Delphi technique is the same with Delphi method but just in one Cycle and also without exact options for choosing. In this way, we should gathering the information and other data from "Referee Papers" as the "First Round" and then share these results with experts to review and modified them as the "Second Round". In addition, we can choose one or two expert(s) for final review of results.

**Table 1 – Elected Factors from ML process**

| 1 | Nature of Detection Efficiency |
|---|---|
| 2 | Low Theoretical Estimated Event Rate |
| 3 | Non-Predictable Events |
| 4 | Limited available or used Large Telescopes - (Main Survey) |
| 5 | Limited Sky Coverage |
| 6 | Vast Observation Time vs. Small Event Timescale (Short-lived Planetary Deviations) |
| 7 | Lack of efficient fast-response Microlensing follow-up |
| 8 | Past Available Funds |
| 9 | Limited knowledge of real-time point-spread-function (PSF) |
| 10 | Need to Advanced Instruments for improve degree of Precision |
| 11 | Low Number of Employee, Faculties, PostDocs and Grad Students |
| 12 | Fitting and analysis problem |
| 13 | False alarm knowledge and statistics |
| 14 | Less Chance to detection of Nearby Planets |
| 15 | Projects Costs |
| 16 | Lack of Experience and Proficiency to use this Method |
| 17 | There isn't any suitable Searching Strategy |
| 18 | Limitation for Sharing Data with/from another Method(s) |
| 19 | Limited Wavelength Coverage |

I read and skimmed most of papers on ML and RV since 1990 and found 216 researchers on ML and more than 650 researches on RV. From this collection, I reviewed most related papers to ESPs and chose 197 experts.

The experts could evaluate the effectiveness of the factors with grading from 1 (for non-effective factor) to 5 (most effective factor). Also, they could just say the difficulties in the form of explanation or just say their opinion briefly and then I evaluated factors with a quality to quantity converting (Text to Grade).

Unfortunately, less than half of experts answered these questions. But fortunately in Delphi method, as a rule, complexity of the project show the minimum required experts (more complex: more experts), and a good number of experts for this type project is about 5-10 persons and more than 20 experts is ideal.

Even after evaluating the factors, decision about choosing the most effective factors on ML rare results is so hard. Because, averaging the factors' grades isn't a good solution. In many times, you are loosing your valuable data to a global approach to the mean of the grades (e.g. "3"). For solving this problem I used Analytic Hierarchy Process (AHP) method to make a "correct" decision about the factors' effectiveness. In a rapid definition, AHP is a unique way to show our priorities as a sort of numbers (Forman and Gass 2001).

The effectiveness of the factors shows in Figure-2. The experts voted to "**Nature of Detection Efficiency**" as the most important factor for ML rare results. That means ML can't detect ESPs more than this rate, intrinsically. From an expert word: *"It's simply because microlensing events are very rare and hard to detect. Also only a few percent of known ML events result in detectable planets... ...Most ML events to date have been in the galactic bulge where star densities are high. They require lens and source stars to be aligned to better than a milliarcsecond, and for the best events to a few microarcseconds. For a given star in the galactic centre to undergo microlensing, the chance is about one in a million, and for high magnification microlensing it is more like one in $10^8$."*

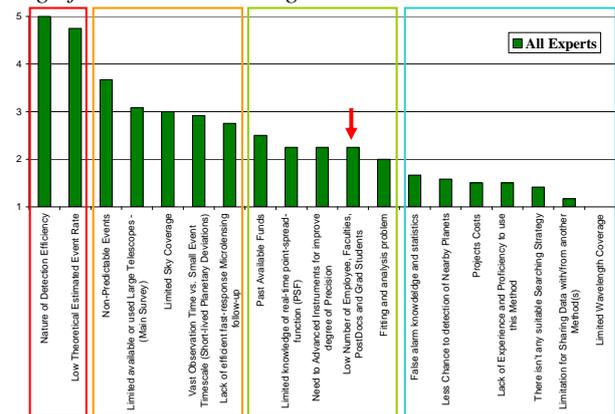

**Figure 2 – Sorted factors from "most effective" to "non-effective"**

This explanation comes from present search strategy and rare results of ML. On the other hand, when we try to solve microlensing events theoretically, we can see the same consequence in the "**Low Theoretical Estimated Event Rate**" form. As you see in Figure-2, experts voted to this factor as the second effective parameter in ML rare results.

Can we trust this evaluation? For answering this question we need another way to calibrating our results at least for one factor.

I chose human resource factor (red arrow in Figure-2) that the experts voted as a low effective parameter on ML results (grade~2.25).

We know from the technology management that technologies lifecycle have the "S" shape curve and with five distinct stages for their lives: Bleeding edge, Leading edge, State of the art, Dated and Obsolete, based on Diffusion of Innovations theory (Rogers 1962). For this curve the X-axis is time and the Y-axis can be the number of papers, patents, trademarks or combination of them. Consequently, the shape of "S" is different for each of these Y-factors and each of them can tell us about the technology life story. Generally, we can detect a new technology when we have a low number of papers and no patent or trademark on that field. High positive slop for papers with low slop of patents and trademarks can show us adult stage of a technology and suddenly decreasing in the number of papers with increasing in the number of patents and trademarks can say about the commercial stage of a technology.

From S-curve of a technology, we can say the present situation of the technology. Also we can use it for comparing "Offer and Demand" or "Public Interest" of two technologies or scientific subjects (Molaverdikhani 2005).

First approach is Google Scholar. Google Scholar can provide about 8,200,000 results[3] that I assumed it as a reservoir for these searches.

If the Human resource factor was low effective, then as a consequence we should NOT have a VERY lower popularity on the ML (number of experts, researchers and grad students that can write articles in ML topic) vs. RV.

Figure-3 shows the results of individual searches on ML and RV methods from 1990 to 2009. The number of articles per year is indicator of "Method's Popularity".

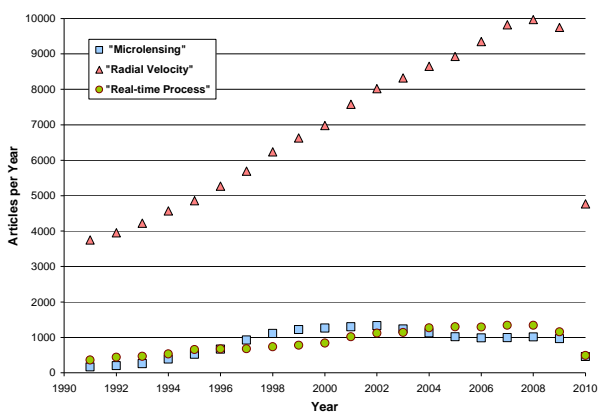

**Figure 3 – Articles in "Microlensing", "Radial Velocity" and "Real-Time Data Processing"**

Obviously, we have a huge different between ML and RV popularities and this is equal to inaccurate votes of experts to "Human Resources" effect.[4]

On the other hand, if we have less human resource in ML but relatively in a long time period, then maybe we can doubt our last result and we can trust to the experts' votes again. Then I used Google Trend (News Archives) as the second approach. News archives can show us a historical background. With some searches on news of Google Trend, I obtain the Figure-4 that it clearly show a strong background of RV vs. ML.

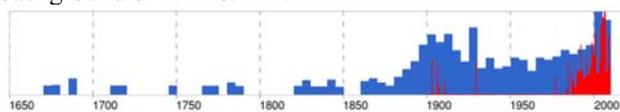

**Figure 4 – News archives for ML (Red) and RV (Blue)**

In this comparison, maximum value of news on each topic normalized to 1 and we just comparing the historical backgrounds. As we expect, because Albert Einstein was born in 1879(!), RV is the older one.

---

[3] Based on December 2009
[4] Because of "Indexing Delay", always we have a negative slop in the last two years, and especially last year (Molaverdikhani 2005).

Therefore, we should find another solution. But, what should we do?

## Categorizing the Experts!

With a new review, I categorized experts in two categories: 1) Special Experts (SE) and 2) Non-Special Experts (NSE). Special experts have at least one direct experience in ESP detection. You can see the fraction of these categories in Figure-5.

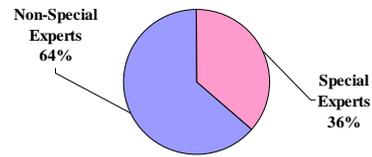

**Figure 5 – Fraction of Special and Non-Special Experts**

Now we can re-evaluate the factors' votes in two separate categories. The Figures 6 and 7 show the results of these evaluations. As you see, human resource has an "Effective" place in the Special Experts' opinion (~3).

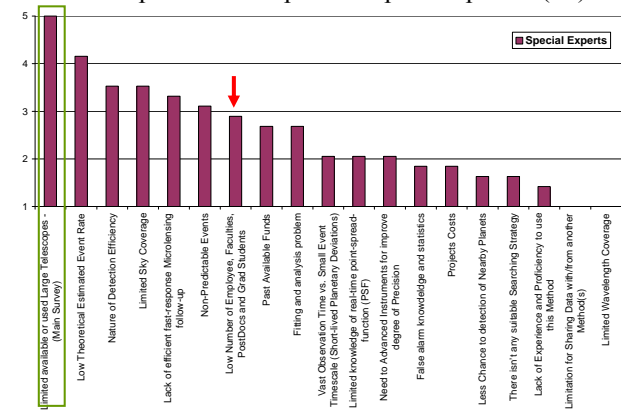

**Figure 6 – Average evaluation of factors in Special group**

But the place of this factor is worse than past, with a "near non-effective" place in the Non-Special Experts' category (~1.6). As a result, I used "Special Expert" group for describing the present situation of ML vs. RV. Also, there are some extreme differences between the results of two categories. You can see the difference of two type experts' opinion in the Figure-8.

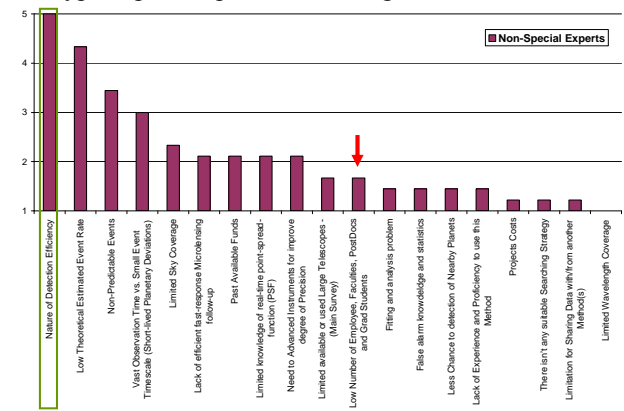

**Figure 7 – Average evaluation of factors in Non-Special group**

Two group opinions are against each other in "Nature of Detection Efficiency" and "Limited Available or Used Large Telescopes" that can use(d) as the main survey telescopes.

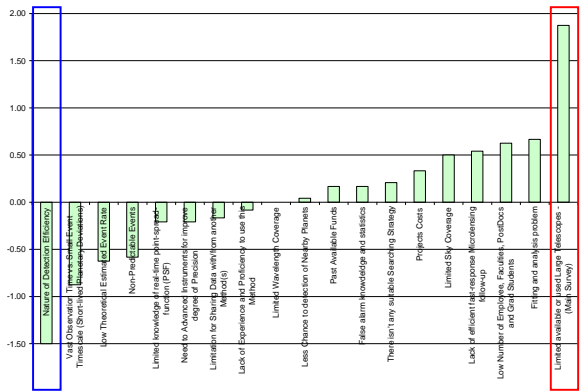

**Figure 8 – Difference of Two group's Opinions**

The NSE group says the nature of ML method is very important and it is one of the main reasons of rare ML results. But the SE group says: "No!" It is not VERY important, It is just important! On the other hand, the NSE says, the number of telescopes for main survey were/are enough to finding a good number of ESPs. But SE again says: "No!!" It is very important and we have NOT enough telescopes for efficient survey."

These differences are very important for ML future that I will return to it in the last part of paper.

## ML present situation vs. RV

Now we know about the correct domain that we can refer to their votes but it is not meaning the same opinion for all experts in this group, which means the MOST of expert VOTED it. In general, if we defining "success" as a more number on detected ESPs then *ML isn't successful*. According to the SE, the most important factors in ML defeat are "**Limited Available or Used Large Telescopes**" and after that "**Low Theoretical Estimated Event Rate**". Second group of important factors are "Nature of Detection Efficiency", "Limited Sky Coverage", "Lack of efficient fast-response Microlensing follow-up", "Non-Predictable Events", "Low Number of Employee, Faculties, PostDocs and Grad Students", "Past Available Funds" and "Fitting and analysis problem".

In a nutshell, the most difficulties in ML come from low **number of telescopes** in main survey and follow-up (Limited availability of very wide field-of-view telescopes spanning the longitudes in the Southern Hemisphere), **nature of method** and **low human and fund resources** (difficulty of the analysis by fitting models to the data and real-time follow-ups problem).

And just as an endorsement, you can see the similar trend for Microlensing and Real-Time Process technology in Figure-3.

### New Members in the Family

On the other hand, new member of planetary systems was born when Bellerophon discovered in 1995; the first hot-Jupiter (Mayor 1995). A giant baby with at least half of Jupiter's mass but just with a 4 days period!

Before this discovery, scientists expected similar planetary systems as we have in the solar system; Cold Gas Giant planets beyond the ice line and Hot, Warm or Cold terrestrial planets (Santos 2005). Now we know about the unexpected hot/warm-Jupiters and also hot/warm Neptunes, Figure-9.

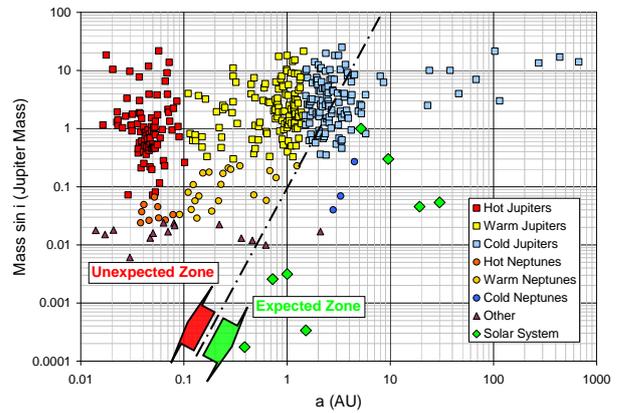

**Figure 9 – New Planets' Types with Expected and Unexpected planets' formation zones before 1995 theories[5]**

As you can see, many of discovered exoplanets are in the unexpected zone (~82%). It was a lucky coincidence for RV surveys and Transit searches that hot and warm Giants exists. If there weren't any hot/warm Giants then we have had just about 76 discovered planets but with more than 12% proportion in the ESPs discoveries for ML (instead present 2% proportion).

The unlucky of ML in finding new planets' types is an important reason of why the microlensing planet count is rather low but, and as we explained, it is not a good excuse for all unsuccessfulness of ML. However, it seems ML is relatively good for solar system type planets detection.

The most important question, after finding difficulties and reasons of unsuccessful results of Microlensing, is: "What is the future of Microlensing?" For answering this question, I have used SWOT method to compare and analyze ML method vs. RV as a success method to finding ESPs.

## Future Strategies from SWOT

The SWOT analysis is often used in a competitive environment between two or more competitors. It is based on identify strengths, weaknesses, opportunities and threats of one competitor against other one. This method is predominantly helpful in identifying areas for future developments.

**Table 2 – ML Strengths and Weaknesses**

| No. | Factor |
|---|---|
| 1 | Direct and accurate measurement of planet Mass |
| 2 | Capable of detecting (with some probability) multiple planets in a single light curve |
| 3 | Detection of Float Planets |
| 4 | More sensitive than most other techniques to small-mass planets |
| 5 | More sensitive than most other techniques to far orbit planets |
| 6 | Only method for far away or extragalactic planet detection |
| 7 | Ability of exomoon Detection |
| 8 | Ability of measurement of planet Orbit (or Period) |
| 9 | Ability of measurement of planet Radii |
| 10 | Chance for detection of Nearby Planets |
| 11 | Non-Predictable Events |
| 12 | Non-Repeatable |
| 13 | Low Event Rate |
| 14 | Disability of measurement extra information (inclination, eccentricity, atmosphere, host star characteristics …) |

In this method, the first step is "definition of objective" for the competition environment. We all know (in public [non-expert] literature) the objective of ESPs detection is equal to successfulness of method and success means "more ESPs". In general, we can explain the strengths and the weaknesses of ML that shown in Table-2.

Figure-10 is shown the **general situation** of these factors in ML vs. RV.

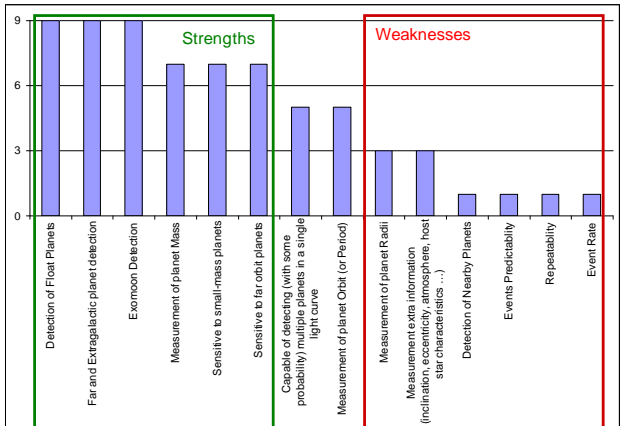

**Figure 10 – General situation of ML's Strengths and Weaknesses**

But how much of these strengths or weaknesses are really helpful for "finding more ESPs"? In the other word, could you find any strength in ML that helps to detection of more ESPs (comparable to RV results)?

According to the official announces of OGLE and MOA (most important projects for the main survey of microlensing events), there are more than 1000 events per year, as shown in Figure-11, but the rate of unrepeated events (in both MOA and OGLE) is about 600-800 per year (about 7000 events since 1998).

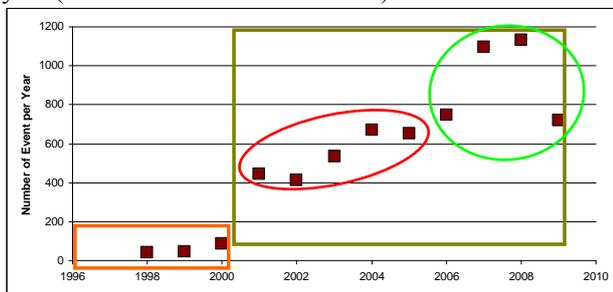

**Figure 11 – Microlensing Events per Year, for OGLE-II, OGLE-III, MOA and MOA-II (Orange rectangle, dark yellow rectangle, red oval and bright green oval)**

And from this rate of events just 5-10% was followed-up. That meaning is in the best case we can follow up 80 events per year (about 500 follow up since 1998). But we just found 10 ESPs (2%) from 500 follow up (0.15% from 7000 events).

Maybe this method has a lot of strengths but any of them is NOT useful for improving the rate of ESPs detection from the present rate, remarkably. Practically, Microlensing is living just in the W-T strategic box (Weaknesses and Threats) and we should turn on the other boxes (W-O, S-T and S-O) for a better future.

### The New Game Board Strategy

We need to change "The Game Board". This is a strategic plan when we don't have any strength in our side. In the new game board strategy, we have rights to change the common game board rules. In our problem, the main rule is "Objective". Then we should change the objective and definition of successfulness of ESPs detection methods. In this way we can use our strengths for definition a new goal. Of course we should have a look to other methods to finding safest plan and also the new objective should have an "acceptable value".

For doing this Game, I used main problems (questions) in planetary science in the field of exoplanets that scientists are trying to answer them. Three fundamental questions are (Lunine 2008):

1. *What are the physical characteristics of planets in the habitable zones around bright, nearby stars?*
2. *What is the architecture of planetary systems?*
3. *When, how and in what environments are planets formed?*

All of these questions are "Valuable" and the new objective can be chosen from each of these questions, combination or all of them. I have chosen all of them for evaluating the ML's strengths and weaknesses.

For finding the "Priority" of the factors in these objectives, again I have used AHP method. The first step of this process is making "paired comparisons" between factors. After that, we should calculate "Normalized Principal Eigen Vector" for each objective matrix. The priority of each factor in each question is shown in Figure-12.

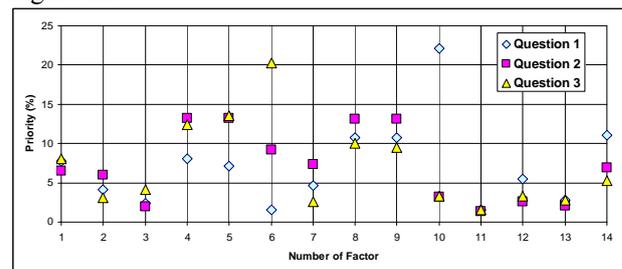

**Figure 12 – The priority of Factors in Fundamental questions**

According to Figure-10, we can discrete general strengths and weaknesses of ML relate to RV. Calculation of "global priorities" is equal to summation of factor's priorities in each category (strengths and weaknesses) in each question. Figure-13 is shown the global priorities. The best objective should have the lowest values in ML weaknesses and the most values in strengths.

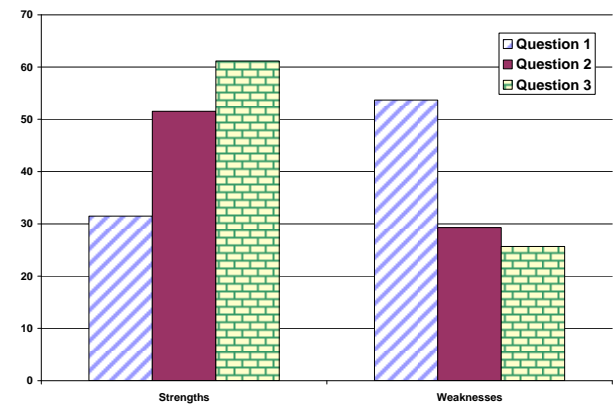

**Figure 13 – Global Priorities for ML in Fundamental Questions**

Obviously "Question 2 and 3" are the best choices for ML, but question 1 is not! The meaning of this result is we should intense our investigation in Question-2 and 3 related topic more than others. The priorities are not constant in the time and it can effect on the results in the future.

There are two ways for solving this problem: the first solution (**Active Solution**) is reviewing this research each five, two or one year; and the second solution (**Passive Solution**) is more investigation on the lower risk objective. A stable objective is low risk ones and it come from more stability in strengths and weaknesses priorities for long term scheduling, relatively. I defined standard deviation of priorities as their stabilities. The range of σ is 2.8 to 7.6 in all questions and factors' categories. Good case should has lower σ in strengths category (means we can trust it in more time that we will have strengths priorities). Figure-14 is shown the relative risk of each objective (2.8-7.6 scaled to 0-100).

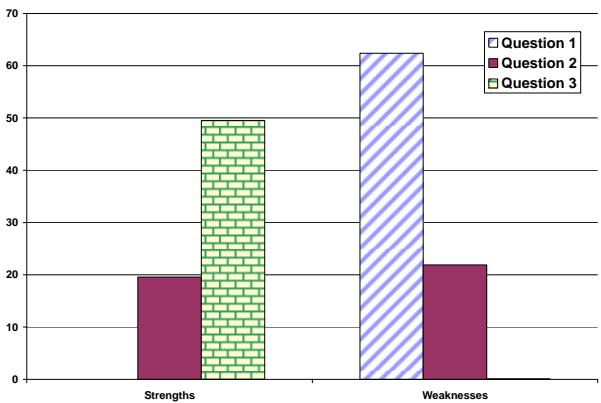

**Figure 14 – Relative Risks of Objectives**

Consequently, I have chosen Question 2 as my main objective because of more stability in its strengths priorities and more strength's global priorities in ML. Microlensing is perfect for solving the architecture of planetary systems problem. With choosing the main objective, we can review our S-W factors toward our objectives, Figure-15 (Strength=Priority×Situation, Weakness=Priority÷Situation).

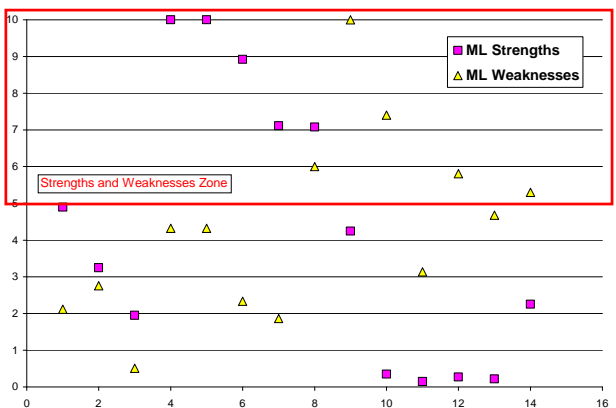

**Figure 15 – S-W factors' scores in objectives (Scaled to 10)**

New Strengths are factors 4, 5, 6 and 7 and new Weaknesses are factors 9, 10, 12 and 14 (factor 8 removed because of equal scores in both regions).

Now, we know about the objectives and situation of ML's strengths and weaknesses. The next step is finding O-T (Opportunities and Threats) for analyzing them with SWOT method.

## Finding Os and Ts!

In the strategic management, each opportunity can play role as a threat and vice versa. This is upon of our decision about using or loosing them. Also, we know, Strengths/Weaknesses are attributes of **Internal Factors** that are helpful/harmful to achieving the objective. On the other hand, Opportunities/Threats are attributes of **External Conditions** that are helpful/harmful to achieving the objective.

In the contrast of a commercial company or business project, a pure scientific project is just based on governmental or quasi-governmental investigations. Investigators have been trying to find the best option with more and effective output and a project can be shutdown easily if they can't show the results.

Best place for finding future investigations is whitepapers, roadmaps and financial reports of related governmental organizations such as NASA and ESA for exoplanets' field. These reports and papers influenced by "all experts" opinions and it is come from "wider communities" of professionals. As a straight result, addition to these resources of information, we should return to our **all expert** database, too.

The most important driver for projects is money and you can see the budget variation of two main supporters of ESPs' detection projects, in Figure-16 (NASA and ESA financial reports).

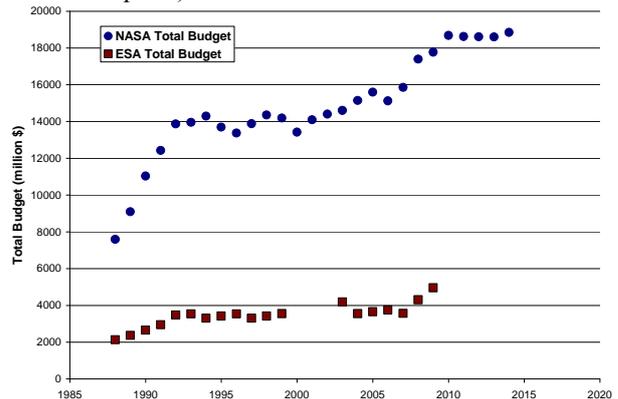

**Figure 16 – NASA and ESA Total Budgets**

It seems the scientists and researches have a better chance in NASA's investments to find their opportunities, but maybe it is not true. I defined the percentage of science budgets from the total budgets as the *"Interest of Agencies"* to science projects; Figure-17 (NASA FY 2010 Budget Request Summary and ESA annually Budget reports).

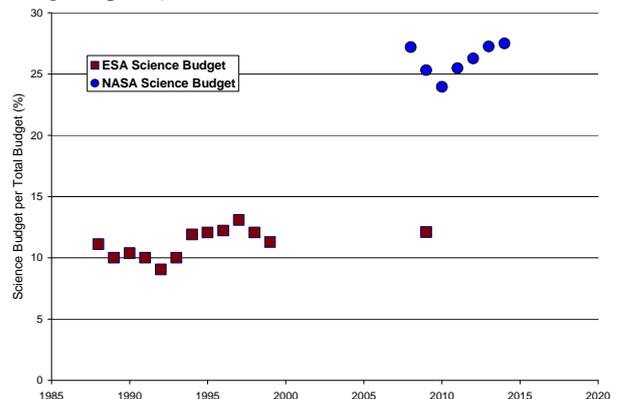

**Figure 17 – Interest of NASA and ESA to Science Projects**

Obviously, NASA is a good choice (but not unique or individual (Gaudi 2009, Dominik 2009)) for finding a fundamental investment for development of a methodology with more budget and higher interest to science projects.

In addition, Microlensing can be referred as an astrophysical method and also as a planetary science one. According to the "science questions and research objective of science plan" for NASA's Science Mission Directorate (2007–2016), planetary theme is about "advance scientific knowledge of the origin and history of the solar system, the potential for life elsewhere, and the hazards and resources present as humans explore space", and astrophysics theme is about "discover the origin, structure, evolution, and destiny of the universe, and search for Earth-like planets".

Then, one of the best opportunities is using both budgets with focusing in the future of their budgets. In Figure-18 you can see the present and future NASA budgets in these fields.

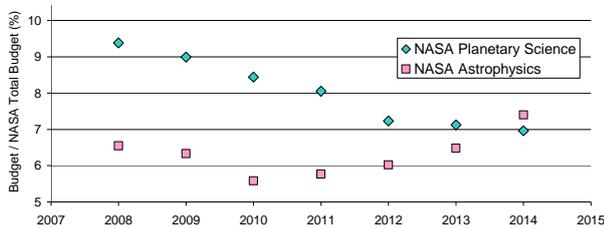

**Figure 18 – NASA's Planetary Science vs. Astrophysics Budgets**

It can help us to find both short term and long term budgets for our planning. Essentially, we should know more about NASA EA (Enterprise Architecture) program. But briefly, it is going to change NASA's Lines of Business (LoB) and it is need to do some infrastructure transformation projects.

This program is very important, because for complement of required infrastructure, NASA is changing its point of view to the "Priority and Rank of Missions". A simple example of these changes is the Space Interferometry Mission, called SIM Lite (or SIM PlanetQuest) with surveying about characterizing other planetary systems and the mission class is "Strategic Large Mission".

This mission endorsed by the 1991 decadal survey as a new Moderate Program but re-endorsed in 2001 decadal survey, and we know it won't launch as soon as 2015. This change in priority is a result of NASA transformation[6].

## Conclusion and SWOT Landscape

In the nutshell, opportunities and threats come from two drivers: **Finance** and **General Opinion**. I explained the finance effects, but general opinion means *"Differences between Special and Non-Special Experts Opinions"*. It can effect in global decisions, easily. In the other word, we need more supports for our projects that to be chosen in a competitive environment of next ESPs' missions election.

Current recommended strategy for finding ESPs wrote for next 15 years (ESPs Committee 2008) and it is acceptable as a present main road strategy. It is included about 20 space and ground missions with RV, ML, Direct (DR), Astrometry (AS) and Transit (TR) methods that you see in Figure-19 and 20 for each method separately (0.5 means a shared mission).

---

[6] NASA Transition Strategy, Version #2.1, February 2008

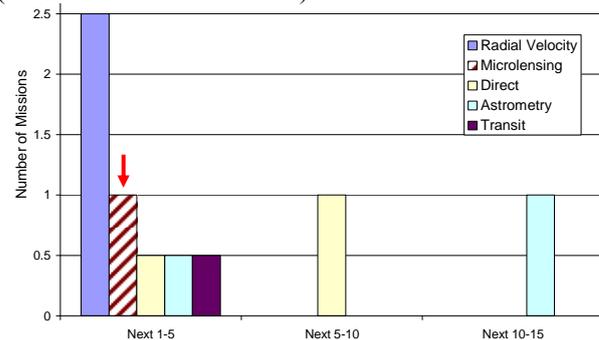

**Figure 19 – Recommended Ground-Based Missions**

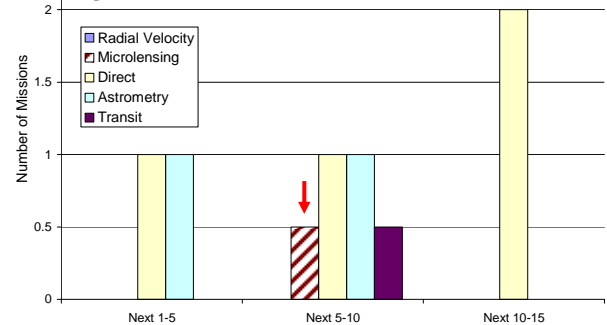

**Figure 20 – Recommended Space-Based Missions**

As you see, in both Space and Ground-based missions, ML has good opportunities and of course some threats. From these recommendations, Opportunities and Threats listed and scored in Table-3. Scores are calculated from differentiation of number of ML missions from number of another technique missions. For example, in ground based missions, RV has 2.5 missions in total and ML has just 1. The differentiation of numbers is 1.5 that means the RV technique has 1.5 (4.3 when scaled to 10) missions more than ML and it is a potential threat for development of ML (T1).

**Table 3 – Opportunities and Threats from future missions**

| | Opportunities | score | Threats | score |
|---|---|---|---|---|
| 1 | TR - G-based missions | 10 | RV - G-based missions | 4.3 |
| 2 | RV - S-based missions | 8 | DR - G-based missions | 1.4 |
| 3 | | | AS - G-based missions | 1.4 |
| 4 | | | DR - S-based missions | 10 |
| 5 | | | AS - S-based missions | 4.3 |

Another example in opportunity zone, there isn't any space based RV but there is a shared recommend mission for ML. This is a potential opportunity for ML is space based missions (O2). Scores scaled to 10 with a little correction on RV space based opportunity. Because of recent investigations of TR and lower probability for next investments, opportunity in TR is more powerful than RV. Now we can explain four SWOT strategies and analyze them based on financial statements and all experts' opinions. Figure-21 (from Appendix-I) is shown SWOT landscape with product of S-W grades by O-T scores.

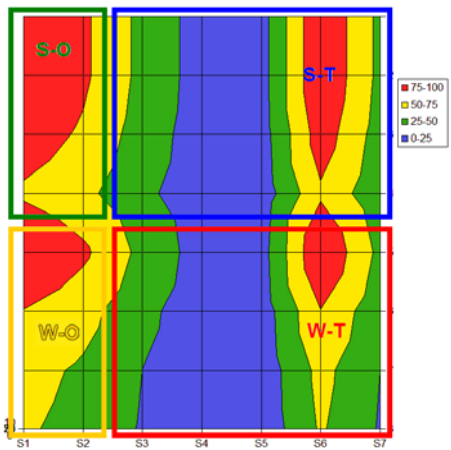

**Figure 21 – SWOT landscape and Four Strategic Zones**

In general, if we have some strength in an opportunity we must use this golden time.

In S-T zone, strategy is defensive and we should minimize our threats effects (although, the next step is an attack for changing threats to opportunities).

In W-O zone, we are loosing our golden time, and we should find some solutions for improve our weaknesses in the opportunities.

And if we have some weakness in the threat region, it is a clear sign of future dangerous events and we must change it to other zones, such as S-T with improving our weaknesses to strengths.

## Suggestions

There are some of briefed and highlighted strategies for Microlensing method in the competitive environment of ESPs' detection.

### S-O

This is the best chance for ML to going forward, as soon as possible. ML is more sensitive than other methods to small-mass planets and other small bodies and also for far orbit objects around their host star (snow line or farther). The lack of ground-based projects for TR method (I think because of Kepler mission effects), is the best chance to improving ML ground-based project everywhere; especially we can target the investigations near TR objectives and far RV objectives (for threats side effects of RV). Allocating smaller class observatories, expanding and sharing ML network with some experienced amateur astronomers can help ML development and expansion. Of course, complement of new generation network by 2014 (Dominik 2009) is another good strategy in this zone of SWOT landscape.

### W-O

Disability of ML to direct measuring of objects' radii is one of the most important weaknesses of ML. We should focused on our other similar strengths (good measurement of objects' mass, etc) and their importance for making a space for entry ML to next 5 years step of ground-based investigations and next 5 to 10 for entry to space missions (as a joint mission with related topics, such as a Joint Dark Energy Mission (NASA ESPs community 2008)). Also another difficulty is disability of finding nearby planets. As the same way, we should focus about ML strengths to detection farthest planets in the Milky Way and extra it.

### S-T

Space missions with Direct method and after that with Astrometry, can ruin the space future of ML. Hopefully, ML has good strengths for small and far bodies in long ways of Galaxy. These can decrease the effect of DR and AS space missions but not for very long time. Speaking about strengths just can stop the rapid growth of other methods and always a space mission for a method is the role of a booster. There isn't any ML space mission for ML at least until 2017. Then ML supporters and experts must try to show the "long term abilities" of ML for measuring properties of planetary systems (focus on Question 2 objective). It can change the threats of other projects (in Short and Middle periods) to a wonderful opportunity (in Long term).

### W-T

The worst zone is W-T. All of threats try to show ML weaknesses. It is very dangerous for future of ML in ESPs detection field. According to conditions, ML supporters must decide about some extra bonus for saving ML. Using space telescopes as the follow-up networks, investment on real-time follow-ups, modeling complex events with more than two lenses, increasing the "popularity" of ML in the universities and using potential financial opportunities in "planetary science" fields are some of suggested extra helps that can save ML future.

## Summary

In this paper, we are looking for reasons of the low results of Microlensing as a contrast with Radial Velocity method and trying to find best strategies for ML success in most possible future of extrasolar planet detection.

In this way, experts voted and explained about selected **effective factors** on ML results and success. Incompatibility of survey results made a force to categorizing experts to *Special* and *Non-special* groups and Specials' votes were closer to "technologies lifecycle trend". Therefore, the most effective difficulty for low results of ML is "Limited Available or Used Large Telescopes" in main survey from Special experts results. But also, detection of new Hot/Warm-Giant planets is the unlucky story of ML.

*Microlensing isn't a successful method for finding more exoplanets (if abundance estimation of hot/warm giants will be a trend of current situation)*, and then we should change our definition of success. In the **new game board strategy**, we can change even our objective. Consequently, I changed ML's objective as the present fundamental questions in planetary science. Probability of success in "architecture of planetary systems" for ML is higher than other methods and I chose it as ML's main objective.

I listed and scored **strengths** and **weaknesses** of ML according to SE group that affected by the new objective. Also, opportunities and threats come from **financial issues** and general opinion about **future missions** (in the form of financial papers, roadmaps, white papers, etc).

There are four main strategies for ML's success in the new objective. We can find the present recommendations and efforts of ML experts for development of ML method in some parts of these suggested strategies. Looking to all possible strategies is **necessary** and you can define them from combination of SWOT factors, but the **Four Critical Strategies** for ML success are:

- o Allocating smaller class observatories; expanding and sharing ML network with some experienced amateur astronomers with the pose of "Only method for finding far-small bodies" [Ground based expansion of ML in next 5 years].
- o Contract a joint space mission (even with a lower proportion percentage on spacecraft opportunities for ML) [Space based expansion of ML in the next 10 years].
- o Showing the strengths of ML in next decadal survey for recommending an individual ML space mission [Long term Space based expansion]. The lack of other space missions in next 20 years would be a golden chance for ML.
- o Investment on expanding the popularity of ML in the universities and students (future experts) in the way of improving infrastructures with using potential financial opportunities in both planetary science and astrophysics fields [making new strengths and opportunities].

I emphasize these four strategies as the main road of future decisions for development of Microlensing method.

### For continuing this work

It is necessary to use PEST analysis to find accurate questions and more possibilities to finding RV and ML difficulties and after that applying Delphi method instead semi-Delphi. Also, you can use raw table on Appendix-II for apply your votes on this systematic survey. Please send it to the author's email and feel free if you have any suggestions or questions.

## Acknowledgement

Author gratefully acknowledges the generous assistance and valuable information provided to me by all experts especially Dr Martin Dominik and Professor Larry W. Esposito.

## Bibliography

Cesarsky C. (ESO) and Canete A.G. (ESA), *"Report No.1 Extra Solar Planets"* ESA-ESO Working Groups, March 2005

Dominik, *"Studying planet populations by gravitational microlensing"*, 2009, unpublished

Dominik, M. 19 other Colleagues, *"Inferring Statistics of Planet Populations by Means of Automated Microlensing Searches"*, 2008, White paper submitted to ESA's Exo-Planet Roadmap Advisory Team

Ergenzinger K., Johann U., Sein E., Stuttard M. and Wallner O.,*"Methods, Technologies and Roadmaps towards Exo-Planet Detection and Characterization from a Prime Contractor's View"*, 2008, White paper submitted to ESA's Exo-Planet Roadmap Advisory Team

Forman E.H. and Gass S.I., *"The Analytic Hierarchy Process: An Exposition"*, Operations Research, Vol. 49, No. 4, July-August 2001, pp. 469-486, DOI: 10.1287/opre.49.4.469.112312001

Gaudi S. and 8 other Colleagues, *"The Demographics of Extrasolar Planets Beyond the Snow Line with Ground-based Microlensing Surveys"*, March 2009, White Paper for the Astro2010 PSF Science Frontier Panel

Kalas Paul and 8 other colleagues, *"Optical Images of an Exosolar Planet 25 Light-Years from Earth"*, 2008, Science 28 November 2008: Vol. 322. no. 5906, pp. 1345 – 1348, DOI: 10.1126/science.1166609

Lawson P. R., Traub W. A. and Unwin S. C., *"NASA Exoplanet Community Report"*, March 2009, JPL Publication 09‐3

Lawson P., Unwin S., and Beichman C., *"Precursor Science for the Terrestrial Planet Finder"*, October 2005, JPL Publication 04-014

Lunine J. and 15 other colleagues, *"Final Report of the ExoPlanet Task Force"*, 18 May 2008, NSF and NASA

Mayor, Michael; Queloz, Didier, *"A Jupiter-mass companion to a solar-type star"* 1995, Nature 378 (6555): 355–359. DOI:10.1038/378355a0

Molaverdikhani K. and Tavakoli A., *"Technology management upon the technology life cycle"*, 3rd International Management Conferences in Iran, 2005

O'Neill Ian, *"New Technique Allows Astronomers to Discover Exoplanets in Old Hubble Images"*, Universe Today, Feb 28, 2009

PPARC, *"Road Map to DARWIN and Beyond: A Ten Year Strategy for Exoplanet Research in the UK 2006 – 2015"*, November 2005, Prepared by the PPARC Exoplanet Forum Working Group for the Astronomy Advisory Panel

Pravdo, Steven H.; Shaklan, Stuart B., *"An Ultracool Star's Candidate Planet"*, ApJ V. 700, Issue 1, pp. 623-632 (2009)., DOI: 10.1088/0004-637X/700/1/623

Rogers, E. M. 1962. *"Diffusion of innovations, first edition"* New York: Free Press

Rowe and Wright, *"The Delphi technique as a forecasting tool: issues and analysis"* International Journal of Forecasting, Volume 15, Issue 4, October 1999, pp. 353-375(23)

Santos N.C., Benz W., and Mayor M. *"Extrasolar Planets: Constraints for Planet Formation Models"*, 2005, Science Vol. 310. no. 5746, pp. 251 – 255, DOI: 10.1126/science.110021

Schneider Jean and 19 other Colleagues, *"The Far Future of Exoplanet Direct Characterization"*, 2009, arXiv:0910.0726

**Appendix-I: Table of SWOT Landscape**

| | | Factors' Numbers | Grades | Opportunities | | Threats | | | | |
|---|---|---|---|---|---|---|---|---|---|---|
| | | | | O1 | O2 | T1 | T2 | T3 | T4 | T5 |
| | | | | 10 | 8 | 4.3 | 1.4 | 1.43 | 10 | 4.3 |
| S1 | Strengths | F4 | 10 | 100 | 80 | 42.8 | 14.3 | 14.3 | 100 | 42.9 |
| S2 | | F5 | 10 | 100 | 80 | 42.8 | 14.3 | 14.3 | 100 | 42.9 |
| S3 | | F6 | 9 | 89.3 | 71.4 | 38.3 | 12.8 | 12.8 | 89.3 | 38.3 |
| S4 | | F7 | 7 | 71.1 | 56.9 | 30.5 | 10.2 | 10.2 | 71.1 | 30.5 |
| W1 | Weaknesses | F9 | 10 | 100 | 80 | 42.9 | 14.3 | 14.3 | 100 | 42.3 |
| W2 | | F10 | 7.406 | 74.0 | 59.2 | 31.7 | 10.6 | 10.6 | 74.0 | 31.7 |
| W3 | | F12 | 5.807 | 58.1 | 46.5 | 24.9 | 8.3 | 8.3 | 58.1 | 24.9 |
| W4 | | F14 | 5.3 | 53.0 | 42.4 | 22.7 | 7.6 | 7.6 | 53.0 | 22.7 |

**Appendix-II: Raw Table of effective factors on Microlensing rare results.**

Use grades 1 to 5; "5" for most effective factor(s) on low result and "1" for non-effective factor(s).
Please add other factors, if you think those are important in the ML rather low results.

| Factor | Effectiveness |
|---|---|
| False alarm knowledge and statistics | |
| Fitting and analysis problem | |
| Lack of efficient fast-response for Microlensing follow-up | |
| Lack of Experience and Proficiency to use this Method | |
| Less Chance to detection of Nearby Planets | |
| Limitation for Sharing Data with/from another Method(s) | |
| Limited available or used Large Telescopes - (Main Survey) | |
| Limited knowledge of real-time point-spread-function (PSF) | |
| Limited Sky Coverage | |
| Limited Wavelength Coverage | |
| Low Number of Employee, Faculties, PostDocs and Grad Students | |
| Low Theoretical Estimated Event Rate | |
| Nature of Detection Efficiency | |
| Need to Advanced Instruments for improve degree of Precision | |
| Non-Predictable Events | |
| Past Available Funds | |
| Projects Costs | |
| There isn't any suitable Searching Strategy | |
| Vast Observation Time vs. Small Event Timescale (Short-lived Planetary Deviations) | |

# Abbreviations

| | |
|---|---|
| **AHP** | Analytic Hierarchy Process |
| **AS** | Astrometry technique |
| **DR** | Direct technique |
| **EA** | Enterprise Architecture |
| **ESP** | Extrasolar Planet |
| **LoB** | Lines of Business |
| **ML** | Microlensing technique |
| **MOA** | Microlensing Observations in Astrophysics |
| **NSE** | Non-Special Experts Group |
| **OGLE** | Optical Gravitational Lensing Experiment |
| **PEST** | Political, Economic, Social, and Technological analysis |
| **RAND** | Research ANd Development corporation |
| **RV** | Radial Velocity technique |
| **SE** | Special Experts Group |
| **SIM** | Space Interferometry Mission |
| **SWOT** | Strengths, Weaknesses, Opportunities and Threats |
| **TR** | Transit technique |